\documentstyle[sprocl,pstricks,pst-coil]{article}

\def\beq{\begin{equation}}
\def\eeq{\end{equation}}
\def\Re{{\rm \,Re\,}}
\def\Im{{\rm \,Im\,}}

\def\vec#1{\mbox{\boldmath $#1$}}

\def\simle{\hspace*{0.2em}\raisebox{0.5ex}{$<$}
     \hspace{-0.8em}\raisebox{-0.3em}{$\sim$}\hspace*{0.2em}}

\begin{document}

\rightline{MKPH-T-96-15}
\vskip 1cm

\title{POLARIZATION PHENOMENA IN SMALL-ANGLE PHOTOPRODUCTION
OF $e^+e^-$ PAIRS \\ AND THE GDH SUM RULE}

\author{A.I. L'VOV ~\footnote{Talk given 
at the {\em Workshop on Virtual Compton
      Scattering}, Clermont-Ferrand, June 1996} }
\address{Lebedev Physical Institute, Russian Academy of Sciences,
    Moscow, 117924, Russia}

\author{S. SCOPETTA}
\address{Department of Physics, University of Perugia, via A. Pascoli,
   I-06100 Perugia, Italy\\
   and Institut f\"ur Kernphysik, Universit\"at Mainz, D-55099, Germany}

\author{D. DRECHSEL, S. SCHERER}
\address{Institut f\"ur Kernphysik, Universit\"at Mainz,
    D-55099, Germany}

\maketitle
\abstracts{
We discuss a possibility to measure the spin-dependent part of the
forward \mbox{Compton} scattering amplitude through interference effects
of the Bethe-Heitler and virtual Compton scattering mechanisms
in photoproduction of $e^+e^-$ pairs at small angles.}

\section{Introduction}
In studies of spin-dependent structure functions of nucleons and nuclei
with real and virtual photons, the verification of the 
Gerasimov--Drell--Hearn sum rule is of special interest \cite{dre95}.
The GDH sum rule is based in essence only on the assumption of
spin-independence of high-energy forward Compton scattering and thus provides
a very clean test of the spin dynamics.

The forward Compton scattering amplitude on a spin-1/2 target is described
by two even functions $f_{1,2}$ of the photon energy $\omega$,
\beq
\label{f-forward}
  f = (\vec e'\cdot \vec e)\,f_1(\omega)
  + i\omega\vec\sigma\cdot(\vec e'\times\vec e)\,f_2(\omega),
\eeq
which, at $\omega=0$, are constrained by the low-energy theorem:
\beq
  f_1(0) = -\frac{\alpha Z^2}{M}, \qquad
  f_2(0) = -\frac{\alpha \kappa^2}{2 M^2}.
\eeq
Here $M,eZ,\kappa$ are the mass, electric charge, and anomalous magnetic
moment of the target, and $\alpha=e^2/4\pi \simeq 1/137$.
Within the framework of the Regge pole model, these functions behave like
\beq
  f_1(\omega) \propto \omega^{\alpha_R(0)},  \quad
  f_2(\omega) \propto \omega^{\alpha_R(0)-1} \quad
   \mbox{for} \quad \omega\to\infty,
\eeq
where $\alpha_R(0)$ is the intercept of the leading $t$-channel
Regge exchange contributing to the amplitudes.
With the usual assumption of $\alpha_R(0)\simle 1$, both $f_{1,2}$ satisfy
{\em once}-subtracted dispersion relations.
The optical theorem allows to find the imaginary parts of $f_{1,2}$
in terms of the total photoabsorption cross sections
$\sigma_{1/2}$ and $\sigma_{3/2}$,
where the subscript refers to total helicities 1/2 and 3/2, respectively:
\beq
 \Im f_1=\frac{\omega}{8\pi}(\sigma_{1/2}+\sigma_{3/2})
        =\frac{\omega}{4\pi}\sigma_{\rm tot}, \quad
 \Im f_2=\frac{1}{8\pi}(\sigma_{1/2}-\sigma_{3/2})
        \stackrel{def}{=} \frac{1}{4\pi}\Delta\sigma.
\eeq
Therefore, one can write the dispersion relation for $f_2$ in the form
\beq
\label{DR-f2}
  f_2(\omega) = - \frac{\alpha\kappa^2}{2 M^2}
  + \frac{\omega^2}{2\pi^2} \int_{\omega_{\rm thr}}^\infty
  \frac{\Delta\sigma(\omega')}{{\omega'}^2-\omega^2-i0^+}\,
     \frac{d\omega'}{\omega'}.
\eeq
The GDH sum rule,
\beq
\label{GDH}
  \int_{\omega_{\rm thr}}^\infty
   \Delta\sigma(\omega')\,\frac{d\omega'}{\omega'}
   = - \frac{\pi^2\alpha\kappa^2}{M^2},
\eeq
arises from (\ref{DR-f2}) under the stronger assumption that
$f_2(\omega) \to 0$ for $\omega\to\infty$,
or alternatively $\alpha_R(0) < 1$.
Numerical investigations \cite{kar73,wor92,san95}
of the integral on the l.h.s. of (\ref{GDH}) in the case of the nucleon
reveal a different behavior for the two isospin combinations
of the proton and neutron amplitudes,
\beq
  f_{1,2}^{I=0} =\frac12(f_{1,2}^p+f_{1,2}^n),  \qquad
  f_{1,2}^{I=1} =\frac12(f_{1,2}^p-f_{1,2}^n),
\eeq
which correspond to isoscalar and isovector exchanges in the $t$-channel.
The $I=0$ GDH sum rule seems to work very successfully, thus
supporting the conjecture that the leading isoscalar Regge exchange
(Pomeron) is decoupled from the spin-dependent transitions.
It is very surprising,
however, that the $I=1$ GDH sum rule
seems to be 
violated, despite the fact that all known isovector 
exchanges (such as the $a_2$-meson) have an intercept 
$\alpha_R(0) \simle 0.5$
and hence cannot spoil the assumption of $f_2^{I=1}\to 0$
for $\omega\to\infty$.

This discrepancy stimulated a lot of efforts to explore
the amplitude $f_2$ experimentally.
A straightforward way to study $\Im f_2$ is provided by measuring
the photoproduction cross sections $\sigma_{1/2}$ and $\sigma_{3/2}$
over a wide energy region, using $4\pi$ detectors for the final hadrons.
In the present paper we discuss a possibility to study 
both $\Re f_2$ and $\Im f_2$ by measuring photoproduction of
small-angle $e^+e^-$ pairs which shows a specific azimuthal asymmetry
due to the interference of Bethe--Heitler and virtual Compton 
contributions.
Such a method was already used \cite{alv73} to determine $\Re f_1$ at
the energy of 2.2 GeV. Using polarized particles, one can
determine $f_2$ as well.

\section{Bethe--Heitler and virtual Compton contributions}

Considering the reaction
\beq
\label{reaction}
  \gamma N \to e^+ e^- N'
\eeq
in the lab frame, we denote the 4-momentum and helicity of $e^+$ by
$\varepsilon_1$, $\vec p_1$, $\frac12 h_1$; we use
$\varepsilon_2$, $\vec p_2$, $\frac12 h_2$ for those of $e^-$ and
$\omega$, $\vec k$, $h_\gamma$ for the photon.
Also, we denote the nucleon spin projections
to the beam direction by $\frac12 h_N$, $\frac12 h_N'$.

Aside from the overall azimuth of the final particles, the kinematics
of $e^+e^-$ photoproduction is specified by four variables.
We choose two of them to be the invariant mass of the $e^+e^-$ pair and
the invariant momentum transfer:
\beq
  W^2 = {k'}^2, \quad k' \stackrel{def}{=} p_1+p_2, \quad
  Q^2=-q^2, \quad q \stackrel{def}{=} p_N'-p_N.
\eeq
Furthermore, we introduce
the fractions of the energy carried by $e^+$ and $e^-$,
\beq
  x_1 = \frac{\varepsilon_1}{\omega'}, \quad
  x_2 = \frac{\varepsilon_2}{\omega'}, \quad
  \omega'= \varepsilon_1+\varepsilon_2,
\eeq
the total momentum of the pair, $\vec k' = \vec p_1 + \vec p_2$,
which determine the angles $\theta_{1,2}$ between $\vec p_{1,2}$ and
$\vec k'$ through
\beq
   2|\vec p_i|\, |\vec k'| \cos\theta_i = 2\varepsilon_i \omega' - W^2,
  \quad i=1,2,
\eeq
and denote the azimuthal angles of
$e^+$ and $e^-$, with respect to the direction of $\vec k'$,
by $\phi_1$ and $\phi_2=\pi+\phi_1$,
see Fig.~1.
In terms of these variables, the recoil nucleon has the kinetic energy and
momenta
\beq
  q_0 = \frac{Q^2}{2M}, \quad q_z=q_0 + \frac{W^2+Q^2}{2\omega},
 \quad q^2_\perp = Q^2 + q_0^2 - q_z^2 \ge 0.
\eeq
Hence, the energy and momentum carried by the $e^+e^-$ pair
is $\omega'=\omega-q_0$ and $\vec k' = \vec k -\vec q$, respectively.

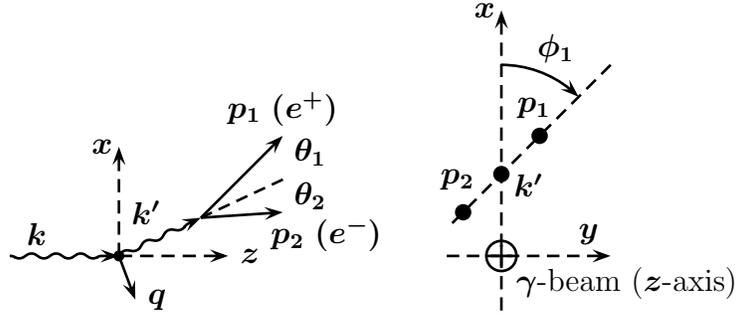
\begin{figure}[hbt]
\large
\begin{center}
\psset{unit=0.4ex,linewidth=1pt}
\begin{pspicture}(0,5)(120,55)
\pszigzag[linewidth=1pt,coilaspect=0,coilwidth=1.5,coilheight=3,
    coilarmA=0,coilarmB=3,linearc=1.5]{->}(00,10)(20,10)
\rput(5,14){\boldmath $k$}
\pscircle*(20,10){1}
\psline[linewidth=1pt,linestyle=dashed]{->}(20,10)(40,10)
\rput(44,10){\boldmath $z$}
\psline[linewidth=1pt,linestyle=dashed]{->}(20,10)(20,30)
\rput(17,30){\boldmath $x$}
\psline[linewidth=1pt,linestyle=solid]{->}(20,10)(23,2)
\rput(27,2){\boldmath $q$}
\pszigzag[linewidth=1pt,coilaspect=0,coilwidth=1.5,coilheight=3,
    coilarmA=0,coilarmB=3,linearc=1.5]{->}(20,10)(35,17)
\psline[linewidth=1pt,linestyle=dashed]{-}(35,17)(50,24)
\rput(25,18){\boldmath $k'$}
\rput(55,29){\boldmath $\theta_1$}
\rput(55,21){\boldmath $\theta_2$}
\psline[linewidth=1pt,linestyle=solid]{->}(35,17)(50,18)
\rput(58,14){\boldmath $p_2$ $(e^-)$}
\psline[linewidth=1pt,linestyle=solid]{->}(35,17)(50,32)
\rput(50,37){\boldmath $p_1$ $(e^+)$}
\pscircle(90,10){3}
\psline[linewidth=1pt,linestyle=solid]{-}(88,10)(92,10)
\psline[linewidth=1pt,linestyle=solid]{-}(90,12)(90,08)
\rput(113,5){\boldmath $\gamma$-beam ($z$-axis)}
\psline[linewidth=1pt,linestyle=dashed]{->}(80,10)(110,10)
\rput(106,14){\boldmath $y$}
\psline[linewidth=1pt,linestyle=dashed]{->}(90,0)(90,55)
\rput(87,55){\boldmath $x$}
\psline[linewidth=1pt,linestyle=dashed]{-}(81,16)(110,45)
\psarcn{->}(90,25){20}{90}{45}
\rput(100,48){\boldmath $\phi_1$}
\pscircle*(97,32){1.5}
\rput(96,37){\boldmath $p_1$}
\pscircle*(83,18){1.5}
\rput(82,24){\boldmath $p_2$}
\pscircle*(90,25){1.5}
\rput(95,23){\boldmath $k'$}
%
\end{pspicture}
\end{center}
\vspace{2ex}
\caption{Kinematics of $\gamma N\to e^+e^- N'$.
 Rear view of the reference frame.}
\end{figure}

The differential cross section of (\ref{reaction}) reads:
\beq
  d\sigma = \frac{1}{(4\pi)^3}\,\frac{\omega'}{|\vec k'|}\,
  \frac{WdW\,QdQ}{\omega^2}\,dx_1\,\frac{d\phi_1}{2\pi} \,|T|^2,
\eeq
where the appropriate sum and average over spins is implied.

To the lowest order in the electromagnetic coupling $e$,
the amplitude $T=T_{\rm BH}+T_{\rm VCS}$ consists of
the Bethe--Heitler and virtual Compton scattering contributions
shown in Fig.~2.
Typically, we consider the kinematical region of
\beq
\label{kin}
 \omega \sim 1 \mbox{~GeV}, \quad
 W\sim 10 \mbox{~MeV}, \quad Q\sim 100 \mbox{~MeV},
 \quad x_1 \sim x_2 \sim 0.5\,,
\eeq
which corresponds to small angles $p_{1\perp}/\varepsilon_1 \ll 1$
and $p_{2\perp}/\varepsilon_2 \ll 1$ of $e^+$ and $e^-$, respectively.

\begin{figure}[hbt]
\large
\begin{center}
\psset{unit=0.6ex,linewidth=1pt}
\begin{pspicture}(0,0)(70,30)
\pszigzag[linewidth=1pt,coilaspect=0,coilwidth=1.5,coilheight=3,
    coilarmA=0,coilarmB=3,linearc=1.5]{->}(00,20)(09,20)
\rput(3,23){\boldmath $k$}
\pscircle*(10,20){1}
\pszigzag[linewidth=1pt,coilaspect=0,coilwidth=1.5,coilheight=3,
    coilarmA=0,coilarmB=3,linearc=1.5]{->}(10,20)(10,11)
\rput(7,14){\boldmath $q$}
\pscircle*(10,10){1}
\psline[linewidth=1pt,linestyle=solid]{->}(11,20.5)(20,25)
\psline[linewidth=1pt,linestyle=solid]{->}(10,20.0)(20,15)
\rput(23,20){\boldmath $e^+e^-$}
\psline[linewidth=2pt,linestyle=solid]{->}(00,05)(10,10)
\psline[linewidth=2pt,linestyle=solid]{->}(10,10)(20,05)
\rput(5,4){\boldmath $N$}
\pszigzag[linewidth=1pt,coilaspect=0,coilwidth=1.5,coilheight=3,
    coilarmA=0,coilarmB=3,linearc=1.5]{->}(40,20)(49,15)
\rput(44,22){\boldmath $k$}
\pscircle*(50,15){1}
\pszigzag[linewidth=1pt,coilaspect=0,coilwidth=1.5,coilheight=3,
    coilarmA=0,coilarmB=3,linearc=1.5]{->}(50,15)(60,20)
\rput(54,21){\boldmath $k'$}
\psline[linewidth=1pt,linestyle=solid]{->}(60,20)(67,23)
\psline[linewidth=1pt,linestyle=solid]{->}(60,20)(67,17)
\rput(73,20){\boldmath $e^+e^-$}
\psline[linewidth=2pt,linestyle=solid]{->}(40,10)(50,15)
\psline[linewidth=2pt,linestyle=solid]{->}(50,15)(60,10)
\end{pspicture}
\end{center}
\caption{Bethe--Heitler and Virtual Compton Scattering amplitudes.}
\end{figure}
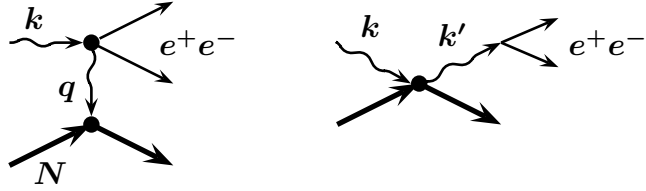

Neglecting the electron mass $m_e$ and keeping leading terms in the
expansion of the matrix elements in powers of the transverse momenta
$p_{1,2\perp}$, we arrive at the following expressions for the amplitudes
diagonal in the spin variables:
\beq
\label{TBH}
 T_{\rm BH} \approx -\frac{2e^3 Z}{Q^2}
  \omega\sqrt{x_1 x_2}\, (\vec e\cdot\vec D) \, (x_1-x_2+h_1 h_\gamma)
  \,\delta_{h_1,-h_2}\,\delta_{h_N,h_N'}
\eeq
and
\beq
\label{TVCS}
 T_{\rm VCS} \approx -\frac{4\pi e}{W^2}
  \sqrt{x_1 x_2}\, (\vec e\cdot\vec d) \, (x_1-x_2+h_1 h_\gamma) \,
  (f_1 - h_\gamma h_N \omega f_2) \,\delta_{h_1,-h_2}\,\delta_{h_N,h_N'}.
\eeq
Here the photon polarization vector $\vec e$
and the vectors $\vec D$, $\vec d$ read
\beq
  \vec e = \frac{1}{\sqrt2}(-h_\gamma \vec e_x - i \vec e_y), \quad
  \vec D = \frac{\vec p_{1\perp}}{p^2_{1\perp}}
         + \frac{\vec p_{2\perp}}{p^2_{2\perp}}, \quad
  \vec d = \frac{\vec p_{1\perp}}{x_1}
         - \frac{\vec p_{2\perp}}{x_2}.
\eeq
The absolute value of $\vec d$ is related to the invariant mass of
the $e^+e^-$ pair, $W^2 \approx x_1 x_2 \vec d^2$.
Another relation, $|\vec D|\approx (x_1 x_2 Q)^{-1}$, is valid provided
$W \ll Q$. We will use these formulae below to discuss
the possibility of measuring $f_2$.

First, a few comments are in order.

(i) Exact calculations of
the amplitudes, accounting for the finite mass $m_e$
and keeping 
higher terms in $p_{1,2\perp}$, agree
to within 1--2\% with the 
approximations
(\ref{TBH}) and (\ref{TVCS})
for all cross sections and
asymmetries considered. 

(ii) Equation (\ref{TBH}) only accounts for the 
contribution of the Coulomb field
to pair production. Magnetic effects due to the spin 
of the target and recoil corrections are of higher order in 
$p_{1,2\perp}$ and were found to be
numerically small and thus irrelevant for our conclusions.

(iii) To obtain Eq. (\ref{TVCS}), we neglected that
the Compton scattering amplitude (\ref{f-forward}) 
varies slowly between $0^o$ and a typical angle of
$q_\perp / \omega \sim 0.1 
\simeq 6^\circ$.  Based on 
dispersion calculations \cite{lvo96} of $\gamma p$ scattering at energies
up to 1 GeV, we estimate this deviation to be less than 1--2\%.  In
general, the momentum transfer $Q^2 \sim 0.01 \mbox{~GeV}^2$ is too small
in comparison with the typical hadronic scale $m_\rho^2 \simeq
0.6\mbox{~GeV}^2$ to visibly change the amplitude $T_{\rm VCS}$ 
found in the forward approximation (\ref{f-forward}).

(iv) To derive Eq.~(\ref{TVCS}), we also neglected the contribution 
of longitudinal photons. Considering a resonance model of $\gamma p$
scattering at energies $\sim 1$ GeV, we have estimated the longitudinal
contribution to $T_{\rm VCS}$ to be of the order of $q_\perp W/\omega^2
\sim 10^{-3}$ with respect to the dominating transverse contribution
(\ref{TVCS}).

\section{Azimuthal asymmetries and determination of $f_2$}

Generally, the VCS amplitude (\ref{TVCS}) is small in comparison with
the BH amplitude (\ref{TBH}) and can be observed only through the
VCS--BH interference.
Note that the $e^+e^-$ pairs produced by a single virtual photon,
like in the VCS process, or in the background reaction
\beq
\label{backgr}
 \gamma p \to \pi^0 X \to e^+e^-\gamma X,
\eeq
have negative intrinsic $C$-parity. Therefore, all the appropriate amplitudes
are even under the interchange
\beq
\label{C}
  \varepsilon_1,\vec p_1,h_1 \leftrightarrow \varepsilon_2,\vec p_2,h_2 \,,
\eeq
while the BH amplitude is odd under (\ref{C}), and its interference
with the VCS contribution leading to the {\em same} final state
results in a $1 \leftrightarrow 2$ asymmetry.

The amplitude (\ref{TBH}) gives the Bethe--Heitler cross section
\beq
\label{csBH}
  \sigma_{12}^{\rm BH} \stackrel{def}{=}
  2\pi WQ \frac{d^4\sigma^{\rm BH}} {dW dQ\,dx_1 d\phi_1}
  \approx \frac{8 \alpha^3 Z^2 W^2}{Q^2}
  \, x_1 x_2\,(x_1^2 + x_2^2) \vec D^2\,,
\eeq
which is symmetric under $1 \leftrightarrow 2$ and depends
on the azimuth $\phi_1$ only weakly.
In the kinematics of (\ref{kin}) it equals 2.4 nb.

The excess of $e^+$ over $e^-$ at some angles
is due to the interference term:
\beq
 \sigma_{12} - \sigma_{21}  \approx
  \frac{16\alpha^2 Z}{\omega} \, x_1 x_2\,(x_1^2 + x_2^2)
  |\vec d|\,|\vec D|\, A(\omega,\phi_1),
\eeq
where the function
\begin{eqnarray}
  A(\omega,\phi_1) &=&
 (\Re f_1(\omega)-h_\gamma h_N \omega \Re f_2(\omega)) \cos\phi_1
  \nonumber \\ && \qquad {} +
 (-h_\gamma \Im f_1(\omega)+ h_N \omega \Im f_2(\omega)) \sin\phi_1
\end{eqnarray}
carries information on the real and imaginary parts of $f_{1,2}$.
This function can be measured through the $e^+$-$e^-$ asymmetry,
\beq
 \Sigma =
 \frac{\sigma_{12} - \sigma_{21}} {\sigma_{12} + \sigma_{21}}
 \simeq \frac{ 2\Re [T_{\rm VCS} T_{\rm BH}^*]} {|T_{\rm BH}|^2}
 \approx \frac{Q^3\sqrt{x_1 x_2}}{\alpha Z\omega W} A(\omega,\phi_1),
\eeq
provided the background contribution such as (\ref{backgr})
is subtracted. The available knowledge on $f_1$
can be used for checking the whole procedure.

It is interesting to note that even with unpolarized photons
one can measure the GDH cross section $\Delta\sigma \propto \Im f_2$.
The appropriate asymmetry
\beq
 \frac12(\Sigma(h_N)-\Sigma(-h_N)) \approx h_N\sin\phi_1\,
 \frac{Q^3\sqrt{x_1 x_2}}{4\pi \alpha Z W} \Delta\sigma
\eeq
is proportional to the target polarization ($h_N$) and
equals 14\% for $\Delta\sigma=100\,\mu$b in the kinematics
of (\ref{kin}), independently of the photon energy $\omega$.
With circularly-polarized photons and a polarized target one
can measure $\Re f_2$ as well and use the dispersion relation
(\ref{DR-f2}) to learn about
$\Delta\sigma$ at asymptotically high energies.
Preliminary estimates of count rates and backgrounds support
a feasibility of such measurements both for nucleons and nuclei.

\section*{Acknowlegdements}
A.L. thanks the theory group of the Institut f\"ur Kernphysik 
and the SFB 201 for their hospitality and support during his stay in Mainz
where this work was done.

\end{document}